# Optical properties of electrostatically assembled films of CdS and ZnS colloid nanoparticles

Suryajaya [(1)], A. Nabok [(1)], F. Davis [(2)], A. Hassan [(1)], S.P.J. Higson [(2)], J. Evans-Freeman [(1)]

[(1)] Materials and Engineering Research Institute, Sheffield Hallam University, City Campus, Pond Street, Sheffield, S1 1WB, UK, phone: +44 114 2253254, e-mail: s.suryajaya@shu.ac.uk
[(2)] Institute of Bioscience and Technology, Cranfield University, Silsoe, Bedfordshire, UK

## Abstract

This work presents a cost-effective alternative technology for the formation of nanostructured semiconductor materials for tunable light emitting devices. CdS and ZnS semiconducting colloid nanoparticles coated with organic shell, containing either $SO_3^-$ or $NH_3^+$ groups, were deposited as thin films using the technique of electrostatic self-assembly. The films produced were characterized with UV-vis spectroscopy, spectroscopic ellipsometry and AFM. UV-vis spectra show a substantial blue shift of the main absorption band of both CdS and ZnS with respect to the bulk materials due to the phenomenon of quantum confinement. The nanoparticles' radius of 1.8 nm, evaluated from the spectral shift, corresponds well to the film thickness obtained by ellipsometry. AFM shows the formation of larger aggregates of nanoparticles on solid surfaces.

Keywords: CdS, ZnS, colloid nanoparticles, electrostatic self-assembly, quantum confinement, ellipsometry, AFM

## 1. INTRODUCTION

In the past decade, II-VI semiconductor nanoparticles attract much attention because of their size-dependent (and thus tunable) photo- and electro- luminescence properties and promising applications in optoelectronics. Different chemical (or wet) techniques for the preparation of II-VI semiconductor nanoparticles in organic and inorganic media, such as LB film method [1,2], Electrochemical Deposition [3,4] and Electrostatic Self-Assembly [5,6], have been exploited. The main purpose of this research is to provide convenient, economic and environmental friendly alternatives to expensive physical methods of nano-structures formation. In this work, we use an aqueous-phase synthesis of colloid nanoparticles' solutions, similar to the route described in [7]. A small modification of the synthesis allowed us to produce CdS and ZnS colloid nanoparticles coated with organic shell containing either $SO_3^-$ or $NH_3^+$ groups and thus suitable for electrostatic self-assembly (ESA). Optical properties of thin films containing II-VI nanoparticles were studied using Varian CARY 50 UV-vis spectrophotometer and J.A. Woollam M2000V spectroscopic ellipsometry; AFM (Nanoscope IIIa) was sued to study films' morphology.

## 2. EXPERIMENTAL DETAILS

### 2.1. Colloid nanoparticles preparation

CdS-$SO_3^-$ and ZnS-$SO_3^-$ colloid nanoparticles were prepared by mixing aqueous solutions of 0.04 M sodium 2-mercapto-ethane-sulfonate with 0.04 M solution of either $CdCl_2$ or $ZnCl_2$. Then 0.04 M solution of sodium sulfide is added very slowly to the solution while it is stirred. Colloid nanoparticles of CdS-$NH_3^+$ and ZnS-$NH_3^+$ are made the same way but using 0.04 M cysteamine hydrochloride as a capping agent.
All chemicals used were of high purity purchased from Sigma-Aldrich. Solutions were prepared using Millipore water having resistance of no less than 18 MΩ.

### 2.2. Electrostatic self-assembly

The films of CdS and ZnS nanoparticles coated with either negatively ($SO_3^-$) or positively ($NH_3^+$) charged shells were deposited layer-by-layer onto electrically charged solid substrates using intermediate layers of either polyanions, such as poly-allylamine hydrochloride (PAH), or polycations, such as poly-styrene-sulfonate (PSS). The substrates (glass and quartz slides, and silicon wafers) were treated in 1% KOH solution in 60% ethanol, to make them negatively charged.
Multilayers of CdS-$SO_3^-$ and ZnS-$SO_3^-$ nanoparticles, were deposited by dipping the substrate consecutively into 1M PAH solution and then into respective colloid solution for 10 minutes in each, as shown in Fig.1.

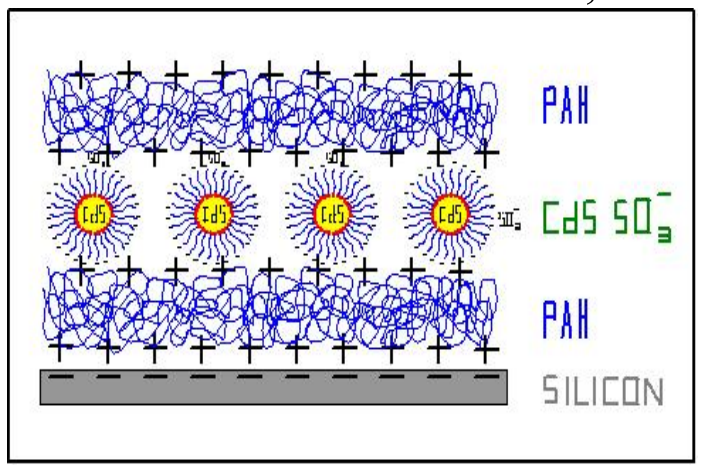

**Fig. 1.** Deposition of CdS-$SO_3^-$ colloid nanoparticles.

Multilayers of $CdS-NH_3^+$ and $ZnS-NH_3^+$ nanoparticles, were prepared by dipping the substrate firstly into 1M solution of PAA for 20 min followed by consecutive dipping in 1M PSS solution and in respective colloid solution for 10 min in each, as shown in Fig.2. All samples were thoroughly rinsed in Millipore water after each deposition step.

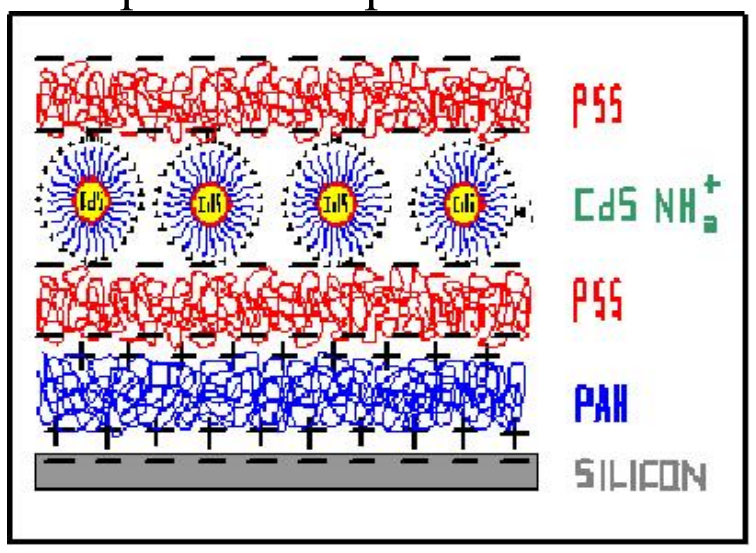

**Fig. 2.** Deposition of $CdS-NH_3^+$ colloids nanoparticles

## 3. RESULTS AND DISCUSSION

Typical absorption spectra of CdS and ZnS nanoparticles embeded in organic films of PAH and PSS are shown in Fig. 3. (a) and (b). The main absorption bands were found at 405 nm for CdS, and 290 nm for ZnS. Both of them are blue shifted from the respective absorption bands of 512 nm and 335 nm for bulk CdS and ZnS materials. This is believed to be due to the effect of quantum confinement, and the size of nanoparticles can be evaluated from the values of blue shift of the absorption bands [1, 8]. Similar spectra were observed for CdS and ZnS nanoparticles capped with $NH_3^+$ groups. In order to obtain the exact positions of absorption maxima, Gaussian fitting of absorption spectra was performed, similar to [9], as shown in Fig. 3 (c) and (d).

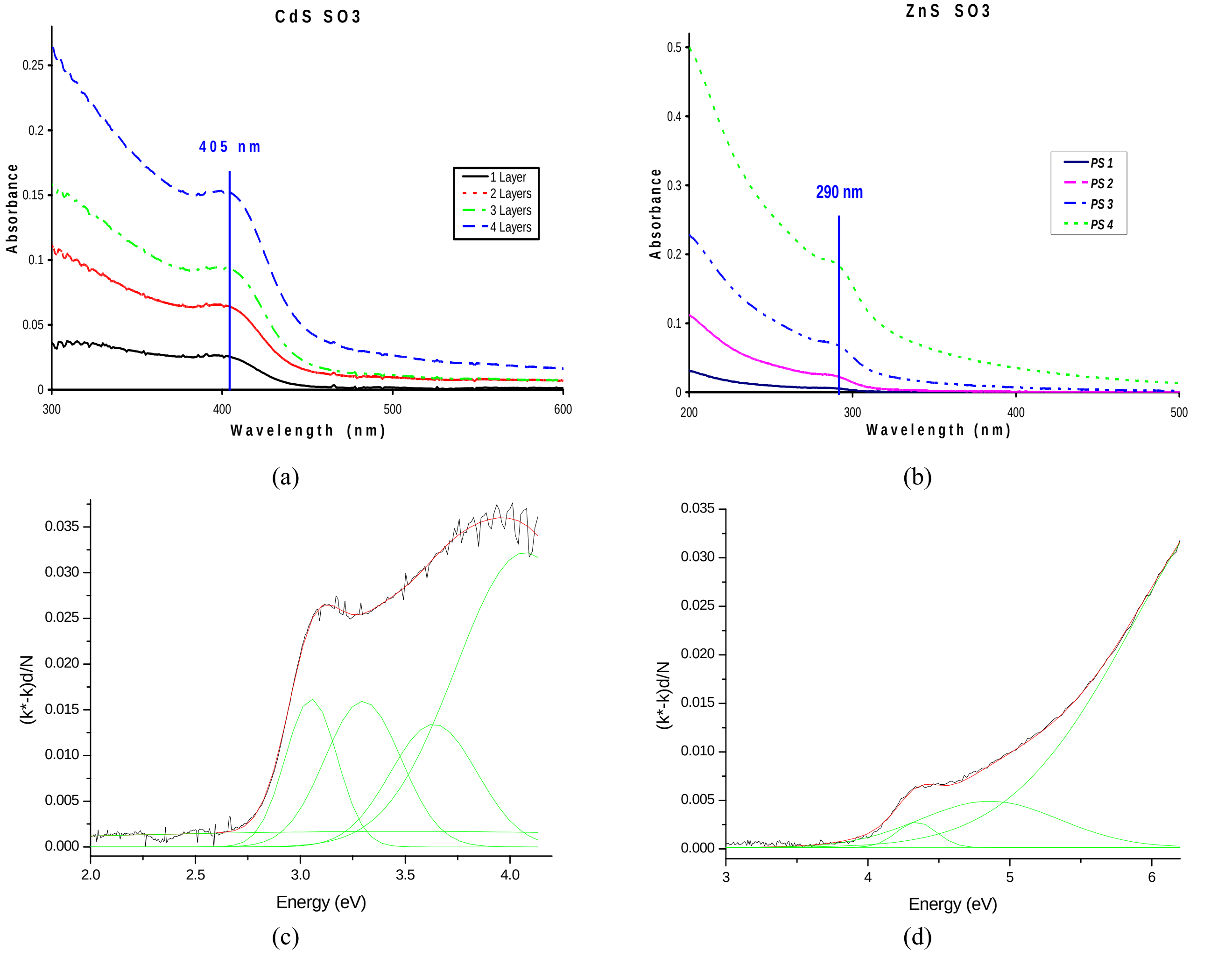

**Fig. 3.** UV–vis absorption spectra of four layers of $CdS-SO_3^-$ **(a)** , four layers of $ZnS-SO_3^-$ **(b)**; Gaussian fitting of the spectra of the first layer of $CdS-SO_3^-$ **(c)** and the first layer of $ZnS-SO_3^-$ **(d)**.

Radius of semiconductor clusters can be calculated using Efros equation [8] for the energy shift, assuming that particles are smaller than Bohr exciton radius.

$$E_{(n,l)} = E_g + \frac{\hbar^2}{2\mu R^2}\phi^2_{(n,l)}.$$

Here $\mu$ is the reduced effective mass of exciton, $\frac{1}{\mu} = \frac{1}{m_e^*} + \frac{1}{m_h^*}$, and $\phi_{(n,l)}$ are routes of Bessel functions (for the ground state $\phi_{(0,1)} = \pi$ ).

The obtained values of particle size are summarized in Table 1.

Table 1
The results of Gaussian fitting of the absorption spectra of CdS and ZnS nanoparticles

| | **E (eV)** | **Radius (nm)** |
|---|---|---|
| $CdS\text{-}SO_3^-$ | 3.12 | 1.8 |
| $CdS\text{-}NH_3^+$ | 3.08 | 1.84 |
| $ZnS\text{-}SO_3^-$ | 4.37 | 1.8 |
| $ZnS\text{-}NH_3^+$ | 4.38 | 1.8 |

Spectroscopic ellipsometry measurements confirmed the growth of polyelectrolyte/nanoparticles films on silicon. For example, the series of $\Delta(\lambda)$ spectra in Fig.4 show their consecutive shift downwards after each layer being deposited, which corresponds to the increase in the film thickness.

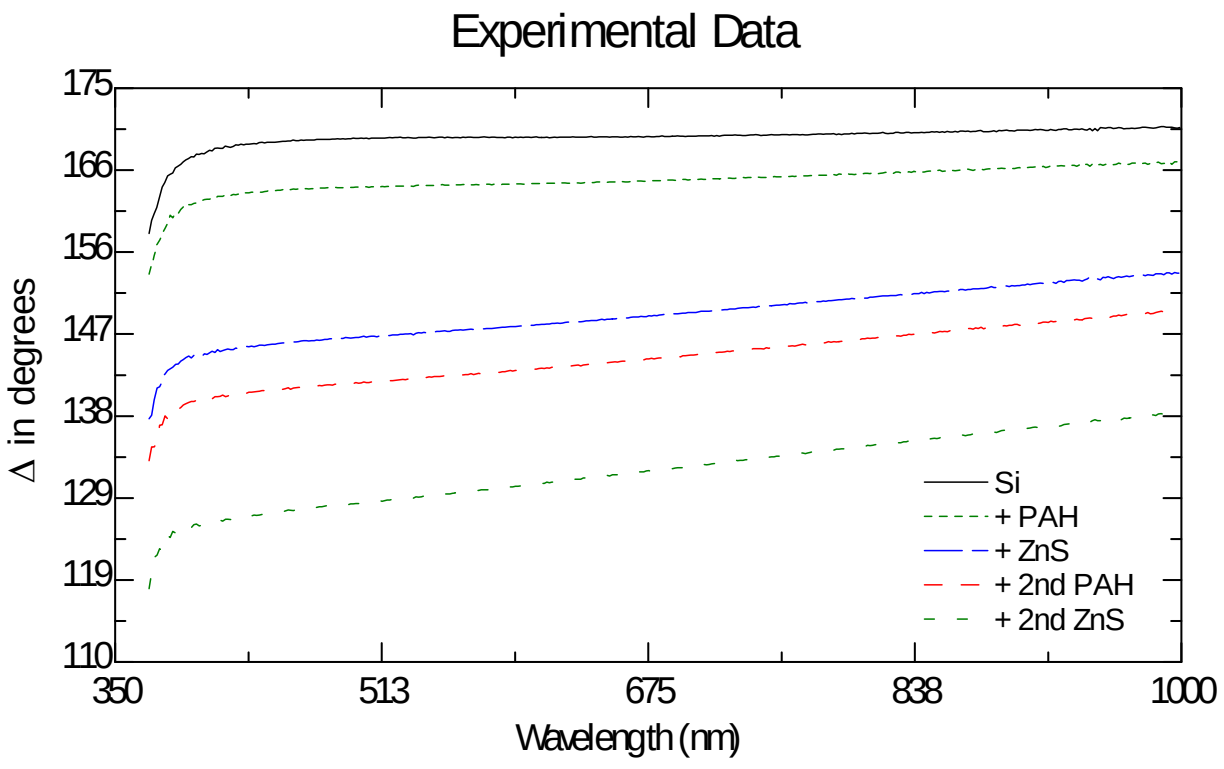

**Fig. 4.** Δ (delta) Ellipsometric spectra

Fitting of experimental Ψ and Δ ellipsometric spectra using VWASE@ J.A. Woollam software allowed the extraction of optical parameters, such as the thickness (d) refractive index (n) and extinction coefficient (k) of all consecutively deposited layers. The results of fitting are summarized in Tables 2 and 3. The obtained thicknesses of nanoparticle layers of around 5 nm for both CdS and ZnS correspond well to size of particles evaluated from UV-vis spectral data considering an additional thickness of organic shells.

Table 2. Ellipsometry fitting of for PAH/CdS-SO3⁻ films

| **Layer** | **d(nm)** | **n (at 633 nm)** | **k (at 633 nm)** |
|---|---|---|---|
| Si | 1mm | 3.867 | 0.02 |
| $SiO_2$ | $5.99 \pm 0.03$ | 1.46 | 0 |
| PAH | $1.91 \pm 0.52$ | 1.49 | 0 |
| 1st CdS | $4.94 \pm 0.04$ | 2.28 | 0.74 |
| PAH | $2.12 \pm 0.02$ | 1.54 | 0 |
| 2nd CdS | $12.84 \pm 0.02$ | 1.82 | 0.74 |

Table 3. Ellipsometry fitting of for PAH/ZnS-SO3⁻ films

| **Layer** | **d(nm)** | **n (at 633 nm)** | **k (at 633 nm)** |
|---|---|---|---|
| Si | 1 mm | 3.867 | 0.02 |
| $SiO_2$ | $3.53 \pm 0.05$ | 1.46 | 0 |
| PAH | $1.96 \pm 0.06$ | 1.49 | 0 |
| 1st ZnS | $5.24 \pm 0.03$ | 2.29 | 0.78 |
| PAH | $2.24 \pm 0.01$ | 1.49 | 0 |
| 2nd ZnS | $5.53 \pm 0.06$ | 2.29 | 0.78 |

Typical tapping mode AFM images of consecutively deposited layers of polyelectrolytes and colloid nanoparticles are shown in Fig. 5 and 6. The first PAH layer in Fig.5 shows non-complete coverage of the Si surface. The next layer of PSS gives much more homogeneous coating. The first layer of $CdS\text{-}SO_3^-$ reveal aggregates of nanoparticles of up to 50 nm in size. Following deposition steps shows an increase in the film roughness and further aggregation of CdS nanoparticles. The aggregation of positively charged CdS nanoparticles can only happen by intercalation with PSS polyanions.

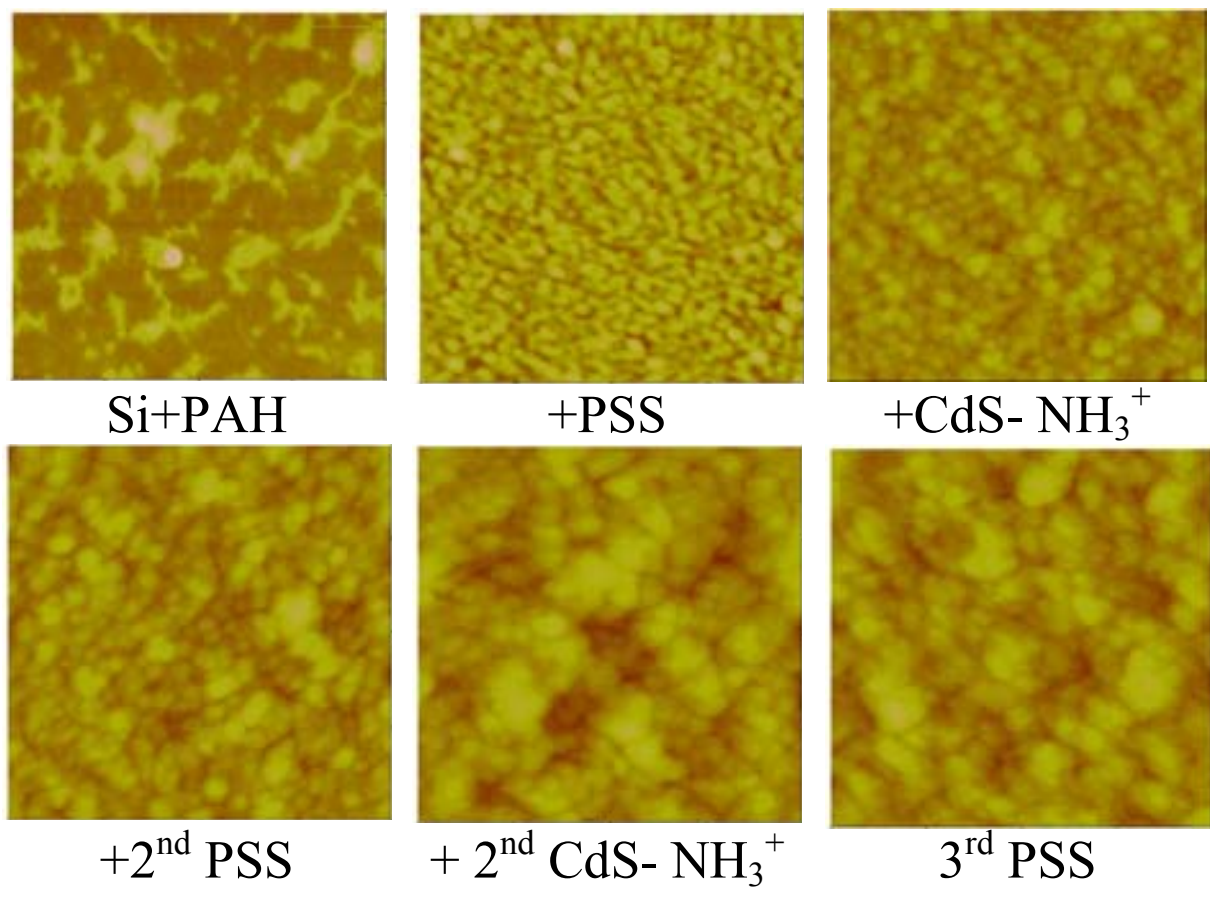

**Fig. 5.** Tapping mode AFM images (1μm in size) of polyelectrolyte/ $CdS\text{-}NH_3^+$ films

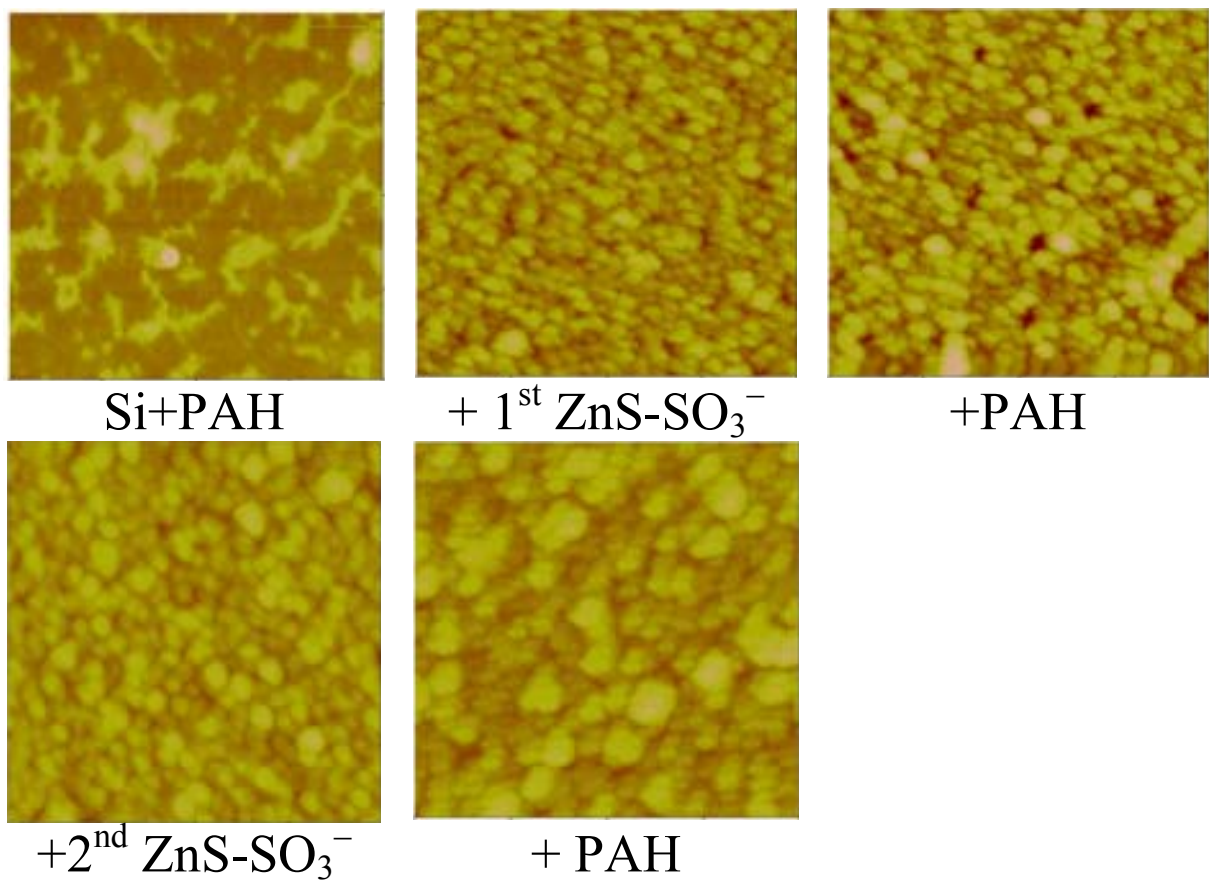

**Fig. 6.** Tapping mode AFM images (1μm in size) of polyelectrolyte/ $ZnS\text{-}SO_3^-$ films

A similar situation is observed for PAH/$ZnS\text{-}NH_3^+$ films. The roughness of layers and aggregation of $ZnS\text{-}SO_3^-$ increases with the number of layers deposited. The size of $ZnS\text{-}SO_3^-$ aggregates is about 40-50 nm. The aggregation of negatively charged CdS nanoparticles can only happen by intercalation with PAH polycations.

The deposition of CdS and ZnS layers from their diluted colloid solutions gives much smaller size of clusters as shown in Fig. 7 (a) and (b).

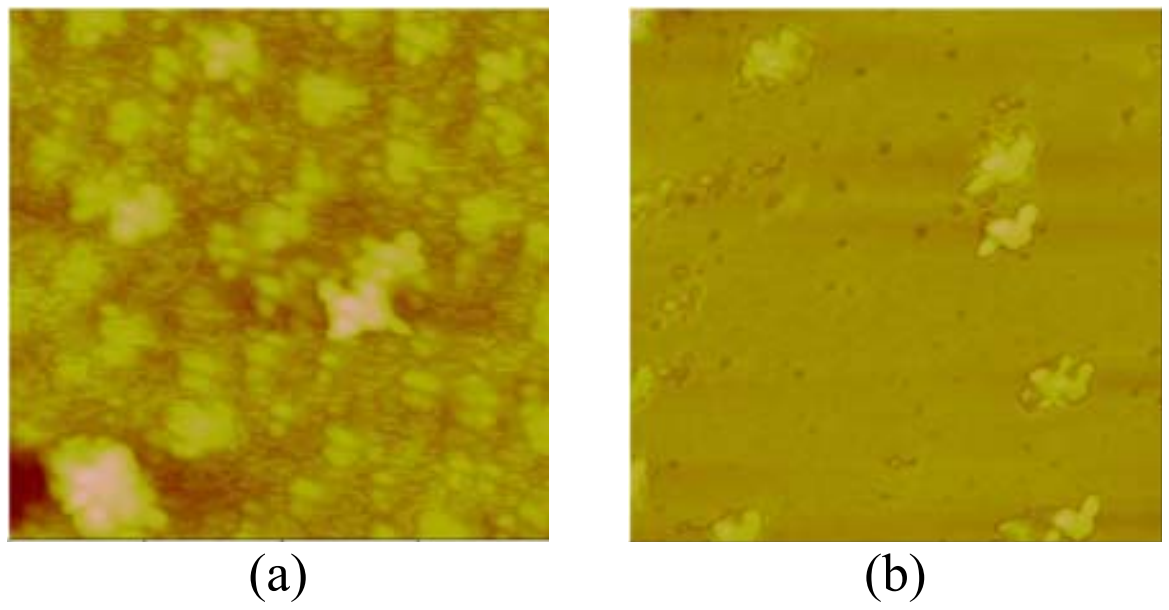

Fig. 7. Tapping mode AFM image (1μm in size) of diluted $CdS\text{-}NH_3^+$ colloid solution, 100 times **(a)** and 500 times **(b)**

## 5. CONCLUSION

Multilayers of CdS and ZnS nanoparticles films were successfully deposited onto quartz and silicon substrate using electrostatic self-assembly via PAH and PSS binding layers. The particle core size of about 2 nm for both CdS and ZnS nanoparticles were obtained from UV-VIS absorption spectra. Ellipsometry gives slightly larger particles' size of about 5 nm due to the contribution of the organic shell. AFM shows large aggregates of electrically charged nano-particles, which forms during their adsorption assisted by the intercalation with oppositely charged polyelectrolytes. The size of aggregates can be decreased from 40-50 nm down to 15-20 nm by diluting colloid solutions.

## REFERENCES

[1] A.V. Nabok, T. Richardson, F. Davis, C.J.M. Stirling, *Langmuir* 13 (1997) 3198.

[2] V. Erokhin, P. Facci, L. Gobbi, S. Dante, *Thin Solid Film* 327-329 (1998) 503

[3] W. Schwarzacher, K. Attenborough, A. Michel, G. Nabiyouni, J. P. Meier, *Journal of Magnetism and Magnetic Materials* 165 (1997) 23

[4] T.L. Wade, R. Vaidyanathan, U. Happek, J.L. Stickney, *Journal of Electroanalytical Chemistry* 500 (2001) 322-332

[5] Y. Lvov, G. Decher, *Crystallography Reports*, vol. 39, 1994, p.628-649.

[6] Janos H. Fendler, *Chem. Mater.* 1996, *8,* 1616-1624.

[7] J. O. Winter, N. Gomez, S. Gatzert, C. E. Schmidt, B. A. Korgel, *Colloid and Surfaces A: Physicichem. Eng. Aspects* 254 (2004) 147-157

[8] A.D. Yoffe, *Advanced in Physics*, 2002, Vol. 51, No 2, 799-890.

[9] A V. Nabok, Richardson T, et. al., (1998), *Thin Solid Films* 327–329 510